\documentclass[prl,showpacs,amssymb,floatfix,twocolumn]{revtex4}
\usepackage{amsmath}
\usepackage{graphicx}
\usepackage{epsfig}
\usepackage{dcolumn}
\usepackage{bm}
\usepackage{times}
\usepackage{epstopdf}

\def\<{\langle}
\def\>{\rangle}
\def\(({\left(}
\def\)){\right)}
\def\[[{\left[}
\def\]]{\right]}

\newcommand{\be}{\begin{equation}}
\newcommand{\ee}{\end{equation}}
\newcommand{\bea}{\begin{eqnarray}}
\newcommand{\eea}{\end{eqnarray}}

\begin{document}

\title{Elusive Spin Glass Phase in the Random Field Ising Model}

\author{Florent Krzakala$^{1,2}$, Federico Ricci-Tersenghi$^{3}$ and
  Lenka Zdeborov\'a$^{2}$}

\affiliation{$^1$ CNRS and ESPCI ParisTech, 10 rue Vauquelin, UMR
  7083 Gulliver, Paris 75000 France\\
  $^2$ Theoretical Division and Center for Nonlinear Studies, Los
  Alamos National Laboratory, NM 87545 USA \\
  $^3$ Dipartimento di Fisica, Sapienza Universit\`a di Roma, INFN -
  Sezione di Roma 1, Statistical Mechanics and Complexity Center (SMC)
  - INFM - CNR, P.le Aldo Moro 2, I-00185 Roma, Italy}

\begin{abstract}
  We consider the random field Ising model and show rigorously that
  the spin glass susceptibility at equilibrium is always bounded by
  the ferromagnetic susceptibility, and therefore that no spin glass
  phase can be present at equilibrium out of the ferromagnetic
  critical line. When the magnetization is, however, fixed to values
  smaller than the equilibrium one, a spin glass phase can exist, as
  we show explicitly on the Bethe lattice.
\end{abstract}
\pacs{75.10.Nr,64.60.De}

\maketitle

Few disordered spin models have generated as much interest and studies
as the Random Field Ising Model (RFIM). Yet, despite four decades of
efforts in mathematics and physics, the thermodynamic properties and
the nature of the phase transitions still remain debated. Originally
proposed by Larkin~\cite{Larkin} for modeling the pinning of vortices
in superconductors, the RFIM has grown to be used for modeling
problems as diverse as (among others) diluted antiferromagnets in a
homogeneous external field \cite{Fishman}, binary liquids in porous
media \cite{DeGennes}, Coulomb ---or electron--- glass \cite{Efros} as
well as systems near the metal-insulator transition \cite{Kirk}. The
non-equilibrium behavior of the RFIM has been used to model the
physics of hysteresis and avalanches \cite{SethnaTarjus} and the model
is also popular in the study of complex systems, for instance to model
opinion dynamics \cite{OPINION}. The Hamiltonian of the RFIM reads
\be
{\cal H} = - \sum_{<ij>} J_{ij} S_i S_j + \sum_i h_i S_i \, ,
\label{Hamitonian}
\ee
where $J_{ij} > 0$ (usually $J_{ij}= 1$), the $N$ Ising spins
$S_i=\pm1$ are placed at the vertices of a graph (usually a periodic
lattice), and the $\{{h_i\}}$ are quenched random fields, usually
having either Gaussian distribution with zero mean and variance
$H_R^2$ or a bimodal distribution $h_i=\pm H_R$.

An important controversy concerning the lower critical dimension has
been resolved using rigorous argument \cite{ImryMa,Proof}, and it is
now known that the RFIM develops long range order for $d>2$. Another
puzzle is associated to the failure of the so-called {\it dimensional
  reduction} property of the RFIM. Standard perturbation theory
predicts to all orders that the critical behavior of the RFIM in
dimension $d$ is the same as that of the pure Ising model in $d-2$
dimensions \cite{ParisiSourlas}, a fact that violates rigorous results
\cite{Proof}. The reason for this failure is often related to the
presence of multiple metastable states; consequently the presence or
the absence of a spin glass (SG) phase in the RFIM has started to
attract a lot of attention.

But is there a thermodynamic SG phase in the RFIM? Based on an
extension of the RFIM to $m$-component vector spins and the large $m$
expansion \cite{intermediate}, it has been argued that in the phase
diagram of the three-dimensional RFIM the paramagnetic and
ferromagnetic phases are separated by a SG phase in which the replica
symmetry is broken, as in mean-field spin glasses~\cite{SpinGlasses}.
Studies using perturbative replica field theory also show the presence
of an intermediate SG phase below six dimensions \cite{CyranoBrezin},
and other claimed that non-perturbative effect would lead to a SG
phase \cite{Dotsenko}. Note that a recent work \cite{Tarjus2} uses
non-perturbative renormalization group to explain the failure of the
dimensional reduction without the use of replica symmetry
breaking. Even in mean field models, the question has sparked debates:
While in the fully connected setting of \cite{SchneiderPytte77} no SG
phase was observed, some works suggested the existence of a such a
phase on the Bethe lattice \cite{BetheYes}, other do not
\cite{BetheNo}, while some remained inconclusive
\cite{BetheBounds}. In numerical studies, altough in \cite{naiveMF}
many solutions to the so-called naive mean-field equations have been
found close to the critical temperature, results from equilibrium
\cite{MC} and out-equilibrium \cite{ParisiRicciRuiz} Monte Carlo
simulations in finite dimension found no evidences for existence of
such a SG phase.

In this paper, we consider this elusive SG phase in RFIM and
show rigorously that the SG susceptibility is always upper-bounded by
the ferromagnetic susceptibility, for any lattice, any dimension and
any choice of fields. Consequently, there cannot be a SG phase out of
the critical ferromagnetic point/line. Secondly, we revise the
solution of the RFIM on the Bethe lattice and show that {\it only}
when the magnetization is fixed to values smaller (in absolute value)
than the equilibrium one a SG phase can exist. 
Note that the RFIM with fixed magnetization appears in many
applications involving a mapping from a lattice gas, e.g.\ in the
Coulomb glasses or in binary liquids.

\paragraph{A rigorous bound on the spin glass susceptibility ---} A commonly accepted definition of a SG phase is the divergence
of the SG susceptibility defined as
\begin{equation}
\chi_{SG} \equiv \frac{1}{N}\sum_{i,j} \Big(\<S_i S_j\>-\<S_i\>\<S_j\>\Big)^2\;,
\label{eq:susc}
\end{equation}
where $\<\cdot\>$ is the thermal average.  The susceptibility
$\chi_{SG}$ is related to the experimentally measured nonlinear
susceptibility \cite{FischerHertz}. In the replica symmetry breaking
theory \cite{SpinGlasses} the celebrated de Almeida-Thouless condition
\cite{deAlmeidaThouless} --- the smallest eigenvalue of the
corresponding Hessian matrix being negative --- implies the divergence
of this $\chi_{SG}$. The study of $\chi_{SG}$ identifies the SG
transition in any theory with (static) replica symmetry breaking
\footnote{In a scenario known as {\it dynamical replica symmetry
    breaking} (see for instance \cite{MontanariSemerjian06}) the spin
  glass susceptibility does not diverge. This scenario is, however,
  impossible in finite dimensional systems. We also checked that it
  does not occur for the RFIM on the Bethe lattice. Beside, the
  ``transition'' into this phase is only a topological one in the free
  energy landscape, with no associated singularity in the free
  energy.} as well as in the droplet model \cite{Droplets}.

Let us now consider a RFIM on a fully connected topology with $J_{ij}
\ge 0$, any other topology can be obtained by setting $J_{ij}=0$ for
all pairs of spins which are not nearest neighbors.  We first prove
that, for any value of the external fields and on any given sample,
connected correlation functions for any pair of spins $i,j$ satisfies
\be
\<S_i S_j\>_c \equiv \<S_i S_j\> - \<S_i\>\<S_j\> \ge 0\; .
\label{FKG}
\ee 
In order to do so, we proceed recursively, and show that if this holds
for a system with $N$ spins, then it holds for a system with $N\!+\!1$
spins. When $N=2$, we have straightforwardly $\<S_1S_2\>_c = 8
\sinh{\((2\beta J_{12}\))}/Z^2$ which is indeed non-negative as long
as $J_{12}\!\ge\!0$. Consider now a system with $N$ spins $S_i$ with
$i=1,\ldots,N$ such that $\forall i,j \in [1\ldots N]$ Eq.~(\ref{FKG})
holds for {\it any choice} of the random fields. We add now a new spin
$S_{N+1}$ with couplings $J_{(N+1)i}$ and external magnetic field
$h_{N+1}$. In the $N\!+\!1$ spins system we denote $w_\pm =
\mathbb{P}[S_{N+1}=\pm1]$.

We now express the correlations in the systems of $N+1$ spins in terms
of the correlations in the system of $N$ spins. First, we evaluate
correlations involving the new spin $S_{N+1}$:
\be
\<S_{N+1} S_i\>^{(N+1)}_c = 2 w_+ w_- \Big(\<S_i\>^{(N)}_+ -
\<S_i\>^{(N)}_-\Big)\; , 
\ee
where the averages $\<\cdot\>^{(N)}_{\pm}$ are computed in the $N$
spins system whose Hamiltonian has been changed by the addition of the
term $-\sum_i \pm J_{(N+1)i} S_i$, which is nothing but a change in
the random fields. Given that the external fields in the measure
$\<\cdot\>^{(N)}_-$ are not greater than the corresponding fields in
$\<\cdot\>^{(N)}_+$, and that susceptibilities are non-negative in the
$N$ spins system by assumption, then $\<S_i\>^{(N)}_-$ is not greater
than $\<S_i\>^{(N)}_+$ and so we have $\<S_{N+1} S_i\>^{(N+1)}_c \ge
0$.

The correlation $\<S_i S_j\>^{(N+1)}_c$ (with $i,j\neq\!N\!+\!1$) is
given by
\bea
\nonumber
\<S_i S_j\>^{(N+1)}_c = w_+ \<S_i S_j\>^{(N)}_{c,+} + w_- \<S_i
S_j\>^{(N)}_{c,-} + \\ 
w_+ w_- \Big(\<S_i\>^{(N)}_+ -
\<S_i\>^{(N)}_-\Big)\Big(\<S_j\>^{(N)}_+ - \<S_j\>^{(N)}_-\Big) \, .
\eea
By the initial assumption, correlations $\< S_i S_j \>^{(N)}_{c,\pm}$
are both non-negative and by the argument used above the last term is
also non-negative, and so $\< S_i S_j \>^{(N+1)}_c \ge 0$. This proves
relation (\ref{FKG}). In fact, what we have proven is a just
particular case of the Fortuin, Kasteleyn and Ginibre \cite{FKG}
inequality, well known in mathematical physics.

Our main point is that from Eq.~(\ref{FKG}) directly follows that the
SG susceptibility is upper-bounded by the ferromagnetic one:
\begin{equation}
\chi_{SG} = \frac{1}{N}\sum_{i,j} \<S_i S_j\>_c^2 \le \frac{1}{N}
\sum_{i,j} \<S_i S_j\>_c = \chi_F\, .
\label{eq:suscInequality}
\end{equation}
This is true on any lattice and for any choice of the external fields,
as long as the pairwise interactions are non-negative (and hence not
frustrated). Eq.~(\ref{eq:suscInequality}) implies in particular that
$\chi_{SG}$ can not diverge if $\chi_F$ stays finite.  We can say even
more: The fact that the correlation matrix has all non-negative
elements, $C_{ij}=\<S_i S_j\>_c \ge 0$, implies that among all
possible susceptibilities the ferromagnetic one, $\chi_F$, is always
the largest one. This means that, in order to understand whether any
kind of long range order develops in a RFIM, it is enough to check
whether the ferromagnetic susceptibility is diverging, and this is a
great step of reductionism! The paramagnetic phase is defined by the
non-divergence of $\chi_F$ thus, clearly, there is no SG phase for
$T>T_c$, where $T_c$ is the ferromagnetic critical temperature.  This
statement allows to reject many predictions in the literature: All
scenarii where $\chi_{SG}$ diverges while $\chi_F$ is finite
\cite{intermediate,CyranoBrezin} are ruled out.

In the ferromagnetic phase, $T<T_c$, the $\chi_F$ would diverge
because of the coexistence between the ``up'' and the ``down'' phases,
with magnetization $m^+$ and $m^-$ respectively. Nevertheless, we can
select one of these two states by using proper boundaries conditions
or by adding an infinitesimal field.  In each of these states the
ferromagnetic susceptibility is finite (they are not critical and the
clustering property holds) and therefore the SG susceptibility
is again finite.

The ferromagnetic susceptibility truly diverges only exactly at a
second order critical point, $T_c$, where two new states are generated
from the paramagnetic one. At this point the Hessian, which is the
inverse of the correlation matrix, develops a zero mode whose
eigenvector has all non-negative elements (thanks to $C_{ij} \ge
0$). In other words, the two new states generated by a second order
transition will have different magnetizations.  The susceptibility
$\chi_F$ is thus diverging exactly at $T_c$, but leads only to a
ferromagnetic long-range order below $T_c$.

It seems to us that the only scenario, we are unable to exclude, for
existence of a SG phase is to have a dense set (e.g.\ in $T$) of
ferromagnetic critical points. We have, however, no reason to believe
that such an exotic scenario appears in the RFIM (nor actually in any
other model that we know of) and thus we conclude that there is no SG
transition in the RFIM.

\paragraph{RFIM with fixed magnetization on the Bethe lattice ---}
\label{sec:RS}
In order to go beyond the strong constraints of
Eq.~(\ref{eq:suscInequality}), we now consider a RFIM where the
magnetization $m$ is fixed to an arbitrary value. 
In this case, it is worth considering the
free energy $f(m)$ as a function of the magnetization $m$.

If two states exist with different magnetization, $m^-\!<m^+$, and if
one fixes the magnetization $m\in(m^-,m^+)$, then in any finite
dimension there is a phase separation between the $m^-$ phase and the
$m^+$ phase, with the appearance of (at least) one domain wall.  The
free energy $f(m)$ is thus concave and given by the Maxwell
construction between $m^+$ and $m^-$.

\begin{figure}[t]
\begin{center}
\includegraphics[width=8.5cm]{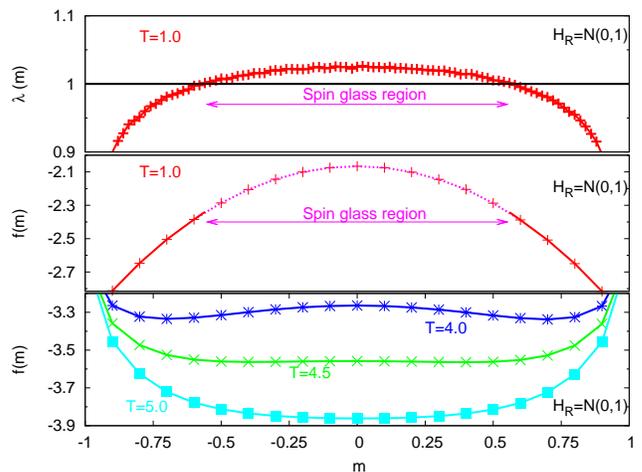}
\end{center}
\caption{(color online) Free energies $f(m)$ and the stability
  parameter $\lambda(m)$ versus the magnetization $m$ on a Bethe
  lattice of coordination $c=6$ with Gaussian random fields of unit
  variance and zero mean. Bottom: we show $T=5$, $T=4.5$ and $T=4.0$
  (with $T_c=4.66$); notice the appearance of two minima for
  $T<T_c$. Middle: a low temperature case ($T=1$). 
A spin glass region
  appears when the magnetization is fixed to low enough values. Top:
  the corresponding stability parameter $\lambda(m)$ showing the spin
  glass order for $|m|<0.56$.\label{fig1}}
\end{figure}

Such arguments, however, do not apply when the RFIM is defined on a
mean-field topology, e.g.\ on the Bethe lattice --- a random graph
with fixed coordination number, $c$. Indeed such random lattices are
expanders, that is the surface-to-volume ratio of any subset of
vertices does not decrease to zero when the subset is made larger (but
still much smaller than the entire lattice). The shape of $f(m)$ thus
does not need to be concave, and indeed we see it develops the
double-well shape typical of mean field ferromagnets for $T\!<\!T_c$
(see Fig.\ref{fig1}). Notice that mean-field geometries are important
in many applications, such as statistical inference and combinatorial
optimization (e.g.\ the RFIM with a fixed magnetization corresponds to
the weighted graph partitioning, a well known NP-hard problem, where
the presence of a SG phase is expected).

Nonetheless the precise determination of $f(m)$ for a given sample or
even for the ensemble average is a nontrivial task.  In principle one
would like to compute the free-energy in the presence of external
field $H_m$ chosen such that the equilibrium magnetization is exactly
$m$ and obtain $f(m) = f(H_m) + H_m m$. However, for a double-well shaped $f(m)$ with minima in $m^-$ and $m^+$, magnetizations in the interval $(m^-,m^+)$ are in principle unreachable, as the Legendre transform computes the convex envelope of the true function $f(m)$.

We now propose an algorithm for computing $f(m)$ on the Bethe lattice
even in these situations.  Our approach is based on the Bethe-Peierls
method, also known as cavity method \cite{MezardParisi} or belief
propagation algorithm \cite{IPC}. For every directed link $(ij)$ we
define a cavity field $u_{i \to j}$ as the effective local magnetic
field which spin $j$ receives from spin $i$.  The cavity fields must
satisfy the following self-consistent equations
\begin{multline}
u_{i \to j} = \frac{1}{\beta} \tanh^{-1}\bigg\{\tanh(\beta J_{ij})\\
\tanh\Big[\beta \Big(H_m + h_i + \sum_{k \in \partial i \setminus j}
u_{k \to i} \Big)\Big]\bigg\}\;,\label{eq:u}
\end{multline}
where the summation is over all neighbors of $i$ but $j$.  The
external uniform field $H_m$ must be chosen such as to fix the global
magnetization to the desired value by
\begin{equation}
m = \frac{1}{N}\sum_i \tanh\bigg[\beta
\Big(H_m + h_i + \sum_{j \in \partial i} u_{j \to i}\Big)\bigg]
\label{eq:m}
\end{equation}
Solved Eqs.~(\ref{eq:u}-\ref{eq:m}), the (extensive) free energy is
given by
\begin{multline}
-\beta F= \sum_{i} \log{\Big\{2\cosh{\big[\beta ( H_m + h_i +
\sum_{j\in \partial i} u_{j\to i})\big]}\Big\}} \\ 
- \sum_{ij} \log \Big\{\sum_{s=\pm 1}e^{s\beta J_{ij}} 2
\cosh[\beta(u_{j\to i}+su_{i\to j})] \Big\}\, .
\end{multline}
In order to fix the magnetization to a value $m$ corresponding to the
non-convex part of the free energy function $f(m)$, we solve
Eqs.~(\ref{eq:u}-\ref{eq:m}) by the following iterative scheme that
forces the procedure to converge to the right fixed point, even when
this is thermodynamically unstable. (A) Set $t\!=\!0$ and assign
random values to $\{u_{i \to j}^{(0)}\}$ and $H_m$.  (B) Repeat (i)
compute $\{u_{i \to j}^{(t)}\}$ by Eq.~(\ref{eq:u}); (ii) compute
$H_m$ solving Eq.~(\ref{eq:m}) by the bisection method; (iii)
increment $t$ by $1$; until a convergence criterion is met or a
maximum number of iterations is reached. (C) If converged, compute the
free energy $f(m)$ using fixed point cavity fields $\{u_{i \to j}^*\}$
and $H_m^*$.

Note that the bound derived in Eq.~(\ref{eq:suscInequality}) is valid
for any external field, and therefore for any values of $m$ such that
$f(m)$ coincides with its convex envelop. This bound is, however,
\emph{not valid} for the values of $m$ in the interval $(m^-,m^+)$ and
this is the place we should check for the appearance of a SG
phase. In order to do so, we study the stability of the Bethe-Peierls
solution towards the appearance of a SG order with a diverging
$\chi_{SG}$.  There are several different methods for computing this
instability (for a review see appendix C in \cite{Thesis}), all of
them generalizing the de Almeida-Thouless condition
\cite{deAlmeidaThouless}. The numerically most precise one is ∂to
study the fate of a small perturbation to the cavity fields
\cite{Pagnani}, that are evolving according to the following linear
equations
\begin{equation}
\delta u_{i \to j}^{(t+1)} =
\frac{\partial u_{i \to j}^{(t+1)}}{\partial u_{k \to i}^{(t)}} \;
\delta u_{k \to i}^{(t)} \;.
\end{equation}
The divergence of the root mean square of the $\delta u$'s signals a
local instability and the appearance of a SG phase. In
practice we measure the parameter $\lambda$ which is the rate of
growth of the root mean square of the $\delta u$'s.

\begin{figure}[t]
\hspace{-0.7cm}
\includegraphics[width=9.25cm]{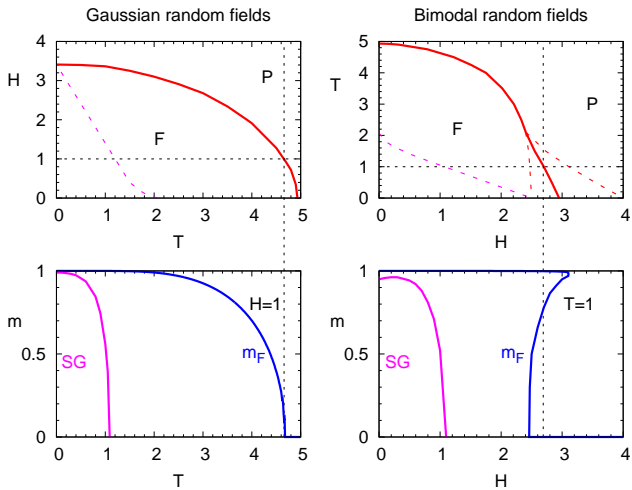}
\caption{(color online) Phase diagram of the RFIM on the Bethe lattice
  with $c=6$ with Gaussian (left) and bimodal (right) random field.
  Top panels: Boundary between paramagnetic (P) and ferromagnetic (F)
  phases (red/full line). In the low $T$ region the transition is
  first order for bimodal fields (red dashed line are the
  spinodals). Below the purple dashed line a spin glass (SG) phase
  exists at zero magnetization $m$. Lower panels: Equilibrium
  magnetization $m_F$ (blue/full line).  A SG phase exist only  
  for $m$ below the purple full line.\label{fig2}}
\end{figure}

In Fig.~\ref{fig1} we show a typical free energy $f(m)$ on a Bethe
lattice and the stability parameter $\lambda$.  We clearly see that
indeed a SG phase is present for some of the non-equilibrium
values of the magnetization.  Fig.~\ref{fig2} shows the phase diagram
of the RFIM on a Bethe lattice.
Note that on a cubic $3d$
lattice with Gaussian random fields, the transition in zero field is
at $T_c\approx 4.5$ and for zero temperature at $H_c\approx
2.3$. Corresponding critical values on the Bethe lattice are larger,
as expected for a mean field approximation. Just as in the mean field
solution of \cite{SchneiderPytte77} the ferromagnetic transition is of
first order for the bimodal distribution of fields at low enough
temperature [for $c\ge 4$, at zero temperature spinodal lines end in
$H_{sp}=(c-1)/2$ and $H_{sf}=c-2$]. The SG region appears
always at a smaller value of magnetization than the equilibrium
one. 
At zero
temperature the ferromagnetic critical point is also critical for the
SG phase with $m=0$. In the renormalization group approach
this is the relevant fixed point and the SG instabilities seen in
the perturbative approach could perhaps be linked to this fact.

\paragraph{Conclusions ---}
We have shown that there is no spin glass phase at equilibrium in the
RFIM, thus closing a long-standing debate on the elusive spin glass
phase in this model. It is \emph{only} if one fixes the magnetization
to non-equilibrium values that a true SG phase can exist, as we showed
explicitly on the Bethe lattice. In finite dimensional systems the
existence such a stable phase, altough unlikely, remains open. This SG
phase, or its vestige, although thermodynamically sub-dominant, may
influence the dynamical behavior of the model. This is particularly
true at zero temperature where the many local energy minima get
stabilized. Our rigorous result also puts a large question mark on the
field theoretical approaches that has lead to erroneous conclusions and
that are still widely used in studies of more complex disordered
systems such a spin glasses.

\end{document}